\documentclass[conference]{IEEEtran}

\usepackage{cite}
\usepackage{amsmath,amssymb,amsfonts}
\usepackage{algorithmic}
\usepackage{graphicx}
\usepackage{textcomp}
\usepackage{xcolor}
\usepackage{subcaption}
\usepackage{multirow}
\usepackage{booktabs}
\AtBeginDocument{%
  }

\begin{document}

\title{Multi-Band mm-Wave Measurement Platform Towards Environment-Aware Beam Management}

\author{\IEEEauthorblockN{Aleksandar Ichkov, Aron Schott, Niklas Beckmann, Ljiljana Simić}
\IEEEauthorblockA{\textit{Institute for Networked Systems, RWTH Aachen University} \\
Kackertstrasse 9, 52072 Aachen, Germany \\
\{aic, asc, nbe, lsi\}@inets.rwth-aachen.de}
}


\pagenumbering{gobble} 




\maketitle

\begin{abstract}
	Agile beam management is key for providing seamless millimeter wave (mm-wave) connectivity given the \mbox{site-specific} spatio-temporal variations of the mm-wave channel. Leveraging non radio frequency (RF) sensor inputs for environment awareness, e.g. via machine learning (ML) techniques, can greatly enhance RF-based beam steering. To overcome the lack of diverse publicly available multi-modal mm-wave datasets for the design and evaluation of such novel beam steering approaches, we demonstrate our software-defined radio \mbox{multi-band} mm-wave measurement platform which integrates multi-modal sensors towards environment-aware beam management.
\end{abstract}

\section{Introduction}

	Providing high rate connectivity in the millimeter-wave (mm-wave) bands is predicated on robust beam steering to overcome the effects of beam misalignment and blockage~\cite{TWC_2021}. However, current state-of-the-art mm-wave beam steering protocols are restricted to radio frequency (RF) based operations with stringent requirements for accurate channel state information (CSI), and are thus susceptible to the spatio-temporal variations of the sparse mm-wave channel, resulting in poor real-world performance due to the overhead/latency constraints of CSI acquisition~\cite{SECON_2022}. To better adapt to the site-specific mm-wave channel, lower complexity out-of-band approaches have been explored, utilizing both RF-based multi-band transmissions and non-RF-based sensor information (see~\cite{Roy_2023} and references therein). In particular, integration of multi-modal sensor inputs such as camera, radar, LiDAR, and GPS, can directly improve RF-based beam steering by \mbox{e. g.} reducing the number of candidate beams during beam sweeping and providing robustness to channel estimation errors. Furthermore, integrating machine learning (ML) can help extract information from the multiple sensor modalities to model the complex channel dynamics and adapt beam steering decisions via environment-awareness in real-time.

\begin{figure*}[t]
	\centering
	\captionsetup{justification=centering}
	\includegraphics[width=0.99\textwidth]{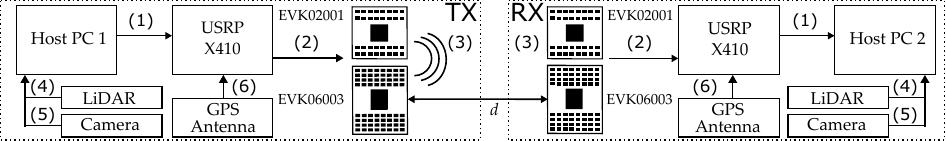}
	\caption{Schematic of the multi-band mm-wave measurement platform with connection types and signals: \\ (1) samples, (2) I--Q baseband signal (bandwidth support up to 400 MHz), (3) RF signal ($f_c$ = 28/60 GHz), \\ (4) 2048$\times$128 point cloud, (5) 360° video stream, (6) 3D position (latitude, longitude and altitude).}
	\label{fig:measurement_setup}
\end{figure*}
	
	Towards this, several research groups are directly working on acquiring multi-modal mm-wave datasets, \mbox{e. g.~\cite{Deepsense6G, FLASH}}. However, the design of practical environment-aware beam steering requires a large number of open, comparable datasets from a multitude of site-specific network deployments in order to train/test ML approaches and most importantly, assess the generalization performance beyond the pre-defined scenario(s). To address this need, we demonstrate our \mbox{multi-band} mm-wave measurement platform which integrates \mbox{multi-modal} sensor inputs, for evaluation of novel beam steering protocols that intelligently adapt to the site-specific channel environment. In contrast to~\cite{Deepsense6G, FLASH}, we implement a packetized directional-TX/directional-RX mm-wave link and several state-of-the-art beam steering protocols, supporting technology-agnostic, wideband, multi-band RF transmissions. Our software defined-radio (SDR) based measurement platform relies solely on commercial-of-the-shelf components, making our system reasonably affordable, readily customizable (in contrast to \mbox{e. g.} FPGA-based platforms~\cite{IMDEA_mmWFPGA}), and easily extendable with additional sensor devices and \mbox{mm-wave} RF \mbox{front-ends}. To support other research groups in leveraging our measurement platform design for generating open, \mbox{multi-modal} mm-wave datasets, we plan to release our GNU Radio framework implementation via~\cite{github_link} as open-source code, once a stable code version is approved.

\section{Multi-Band mm-Wave Beam Steering Support}
	We implement a multi-band mm-wave packetized communication link with beam steering capabilities using a \mbox{8$\times$2 + 8$\times$2} TX/RX phased antenna array evaluation kit EVK02001~\cite{SiversEVK02001} (operating in the 28 GHz band, i.e. 5G-NR FR2-1) and a \mbox{8$\times$4 + 8$\times$4} TX/RX phased antenna array evaluation kit EVK06003~\cite{SiversEVK06003} (operating in the 60 GHz band, \mbox{i. e.} \mbox{5G-NR} FR2-2 and IEEE 802.11ad/ay) by Sivers Semiconductors (\emph{cf.}~Fig.~\ref{fig:measurement_setup}). In principle, our setup allows simultaneous \mbox{multi-band} communication and data collection.
		
	Host PC 1 generates a packet and sends the modulated baseband waveform samples to the SDR (USRP X410), which generates an analog signal with sampling rate $f_s$ at baseband frequency $f_{bb}$ and feeds it to the EVK. The code implementation enables any candidate (standard-compliant) waveform to be streamed via the USRP X410 with bandwidth support of up to 400 MHz. We note that other USRP models are also supported, e.g. X310/N310 but with lower bandwidth support of up to 160/100~MHz. The EVK directly upconverts the baseband signal to the carrier frequency $f_c$ and transmits it over the air. On the RX side, the received signal is downconverted by the EVK, sampled and streamed by the USRP X410 to the host PC 2 where the packet is decoded.

	We implement a number of state-of-the-art beam steering protocols that sweep the full azimuth/elevation angular space by changing the TX/RX beam orientations using a combination of electronic (i.e. 22 beam codebook entries over $[-45^{\circ}, 45^{\circ}]$ azimuth range for both EVK02001/EVK06003) and mechanical beam steering by a custom-made 3D turntable to cover the complete azimuth and elevation range. We can thus collect measurements over fine-grained 3D antenna orientations, estimating the received signal strength and phase and frequency offsets for the respective TX and RX beam entries, which is used for the link establishment and link recovery decision process.

\section{Environment-Awareness via Multi-Modal Sensor Inputs}
	The Ouster OS-1 LiDAR~\cite{OS1} with 128/2048 vertical/horizontal channels (200~m maximum range and 20~Hz frame rate) and a GoPro MAX camera~\cite{GoPro} (360$^\circ$ horizontal field-of-view at \mbox{30 fps}) serve for capturing the 3D environment and object detection. A GPS antenna obtains the 3D position at a rate of 1~Hz. The sensors are connected to the TX/RX host PC, running the Robot Operating System (ROS) in a Docker container~\cite{ika_github_link}, continuously collecting and time-stamping the sensor measurements. To ensure data synchronization for post-processing purposes, the host PC clock is used as a common reference to synchronize the packetized communication link as well as the built-in clock of the GoPro MAX camera. 

\section*{Acknowledgment}
	We thank L. Reiher at the Institute for Automotive Engineering (ika, RWTH) for support with the LiDAR framework. This work has received funding by the German BMBF in the course of the 6GEM research hub, grant number 16KISK038.
\newpage
\bibliographystyle{IEEEtran}
\bibliography{bibliography}


\end{document}